\documentclass[%
superscriptaddress,
twocolumn,
amsmath,amssymb,
showpacs,
prl,
floatfix
]{revtex4}

\usepackage{graphicx}
\usepackage{dcolumn}
\usepackage{bm}

\begin{document}
\title{Observation of zero-field transverse resistance in AlO$_x$/SrTiO$_3$ interface devices }
\author{P.W. Krantz}
\address{Department of Physics, Northwestern University, Evanston, Illinois. 60208, USA}
\author{V. Chandrasekhar}
\address{Department of Physics, Northwestern University, Evanston, Illinois. 60208, USA}
\email{v-chandrasekhar@northwestern.edu}

\date{\today}

\begin{abstract}
Domain walls in AlO$_x$/SrTiO$_3$ (ALO/STO) interface devices at low temperatures give a rise to a new signature in the electrical transport of two-dimensional carrier gases formed at the surfaces or interfaces of STO-based heterostructures:  a finite transverse resistance observed in Hall bars in zero external magnetic field.  This transverse resistance depends on the local domain wall configuration and hence changes with temperature, gate voltage, thermal cycling and position along the sample, and can even change sign as a function of these parameters.   The transverse resistance is observed below $\simeq$ 70 K but grows and changes significantly below $\simeq$40 K, the temperature at which the domain walls become increasingly polar.  Surprisingly, the transverse resistance is much larger in (111) oriented heterostructures in comparison to (001) oriented heterostructures.  Measurements of the capacitance between the conducting interface and an electrode applied to the substrate, which reflect the dielectric constant of the STO, indicate that this difference may be related to the greater variation of the temperature dependent dielectric constant with electric field when the electric field is applied in the [111] direction.  The finite transverse resistance can be explained inhomogeneous current flow due to the preferential transport of current along domain walls that are not collinear with the nominal direction of the injected current.    
\end{abstract}

\maketitle

Two-dimensional carrier gases (2DCGs) in SrTiO$_3$ (STO) based interface devices show a variety of complex correlated electron phenomena, including superconductivity,\cite{reyren,caviglia,benshalom} magnetism\cite{huijben,dikin,bert,li,mehta} and strong spin-orbit interactions.\cite{benshalom,caviglia2,bal}  Almost all of these phenomena are tunable with an electric field, typically applied by means of a voltage $V_g$ applied to the STO substrate so that gating effects are convoluted with the rather complex dielectric properties of STO.\cite{weaver,muller,muller2,rowley}  Recently, it has been recognized that the domain walls formed between tetragonal domains in the low-temperature phase of STO play an important role in determining the properties of 2DCGs in STO based surfaces and interfaces.\cite{chandra, scott,salje,fontcuberta,ma,frenkel,frenkel2,goble,kalisky, christensen}  Here we show that these domain wall give rise to a new effect at low temperatures:  a finite transverse resistance observed in AlO$_x$/SrTiO$_3$ (AO/STO) Hall bars at low temperatures in zero external magnetic field.  The transverse resistance depends on the local domain wall configuration and hence can change with temperature, gate voltage, thermal cycling and position along the sample.   Our results can be explained by inhomogeneous current flow due to the preferential transport of current along domain walls that are not collinear with the nominal direction of the injected current.

STO is a band-gap insulator with a dielectric constant $\epsilon$ of a few hundred at room temperature that rises to a few tens of thousands at low temperatures ($<$20 K).  This increase in $\epsilon$ is associated with an incipient displacive ferroelectricity that is frustrated by quantum fluctuations, the so-called quantum paraelectric transition that occurs around 40 K.\cite{muller,chandra}  STO also undergoes a structural transition from a cubic to tetragonal phase with a slight change of the $c$ axis at 105 K, forming domain walls between tetragonal domains with randomly oriented $c$ axes.  Dielectric spectroscopy shows that the domain walls may be polar,\cite{scott,salje};  local imaging studies have shown that current through 2DCGs flows preferentially along the domain walls\cite{kalisky} and that these domain walls can also be magnetic.\cite{christensen}  As we show below, preferential charge transport along these domain walls gives rise to a finite zero-field transverse resistance (ZFTR) that increases in magnitude significantly below 40 K, near the quantum paraelectric transition of STO.  Its characteristics indicate that it is sensitive to the formation and dynamics of domain walls in STO at low temperatures.

Measurements of the transverse resistance, where the voltage contacts are aligned perpendicular to the path of the current in a Hall bar geometry, are a powerful tool to obtain important information about the intrinsic properties of a material.  In conventional conductors, the transverse resistance is an antisymmetric function of an externally applied magnetic field arising from the Hall effect that depends on the sign, density and mobility of the charge carriers, and as such, vanishes at zero magnetic field.\cite{ziman}  In materials such as ferromagnets with a finite magnetic moment, there might be a component of the transverse resistance that is hysteretic with the applied field and has a non-vanishing contribution at zero applied field.  The origin of this anomalous Hall effect (AHE) arises from broken time-reversal symmetry in the presence of spin-orbit interactions.\cite{nagaosa}  In other cases, the role of time-reversal symmetry breaking is not clear.  For example, inhomogeneous current flow arising from defects\cite{laukhin} or anisotropy in the resistivity tensor\cite{wu} may also give rise to ZFTR without involving time-reversal symmetry breaking.  We propose that a similar explanation describes our observations.    

\begin{figure*}[!ht]
\includegraphics[width=18cm]{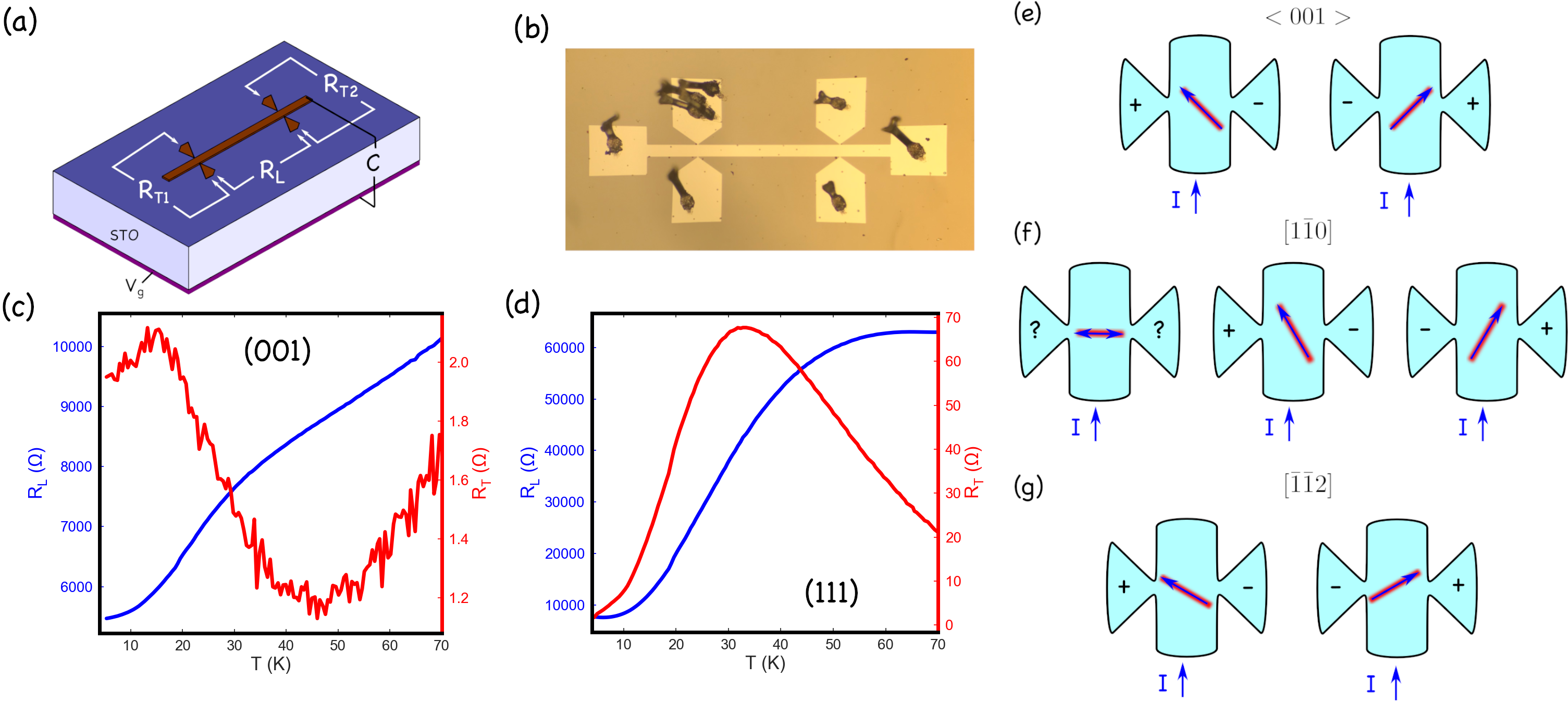}
\caption{ (a) Schematic of a Hall bar, showing the voltage probes used to measure the longitudinal ($R_L$) and transverse ($R_T$) resistances and the capacitance $C$.  The gate voltage $V_g$ is applied to the back of the STO substrate.  (b)  Optical image of a Hall bar.  The width of the Hall bar is 50 $\mu$m and the length between the longitudinal voltage probes is 600 $\mu$m.  (c)  Temperature dependence of $R_L$ (blue) and $R_T$ (red) for a (001) oriented Hall bar.  (d)  Similar data for a (111) Hall bar.   $V_g$=80 V and $B$=0 for data in (c) and (d). (e-g)  Schematic representation of orientation of domain walls near transverse voltage probes.  Red lines represent domain walls, and blue current direction along these domain walls.  (e) [001] oriented sample, with injected current direction along the $<001>$ surface crystal directions.  (f-g) [111] oriented sample with the current injected along the $[1\bar{1}0]$ direction (f) and along the $[\bar{1}\bar{1}2]$ direction (g).  One of the walls in (g) would be oriented along the direction of the injected current and is therefore not shown.}
\label{fig:roft}
\end{figure*}

The samples in our experiments were fabricated by depositing Al in patterned Hall bar configurations on 5 mm x 5 mm Ti-terminated STO substrates to form an amorphous AlO$_x$ layer that pulled oxygen from the STO substrate, resulting in oxygen vacancies and a 2DCG at the AlO$_x$/STO interface.  The Al was deposited in steps of 2 nm, each deposition followed by oxidation in a 100 mT oxygen environment without breaking vacuum.  The process minimizes the strain on the STO surface, resulting in less disorder, as evidenced by the large residual resistance ratios ($\sim$ 10) in our samples.  The Hall bars, each 600 $\mu$m long and 50 $\mu$m wide, were carefully aligned along primary surface crystal directions.  For the (001) STO substrate, these were the equaivalent $<100>$ directions, while for the (111) STO substrate, these were the $[\bar{1}\bar{1}2]$ and $[1\bar{1}0]$ surface crystal directions.  In addition to the longitudinal resistance $R_L$, the sample geometry permitted measurements of two transverse resistances $R_T$ (Figs. \ref{fig:roft}(a) and (b)).   Resistance measurements of the Hall bars and capacitance measurements between the Hall bars and the back gate were performed as a function of temperature $T$, back gate voltage $V_g$ and magnetic field $B$ in a liquid helium cryostat equipped with a superconducting magnet over the temperature range $\sim$5-70 K.   The resistance measurements were performed using lock-in amplifier techniques at low frequency using a custom-built current source.  Capacitance measurements were performed by superposing a 100 mV 1 kHz ac voltage on the dc gate voltage $V_g$, and determining the capacitance from the in-phase and quadrature signals of the current generated in the Hall bar, as described in Ref. [\citenum{davis}].                         

As with other STO-based 2DCGs, the transport properties of the devices can be tuned by an electric field applied by means of a gate voltage to the substrate.  Applying a finite gate voltage $V_g$ at low temperature irreversibly changes the properties of the 2DCG so that the initial resistance at $V_g = 0$ V cannot be recovered unless the sample is warmed up to room temperature before cooling back down.  This phenomenon is well-known in STO-based oxide interfaces.\cite{biscaras}  For this reason, we discuss below only data as $V_g$ was progressively increased from 0 to 80 V.  Sweeping back down in $V_g$ in general results in larger values of resistance at the same value of $V_g$, but the qualitative results do not change. Sheet resistances at $\sim 5$K and $V_g=0$ V were $R_\square \sim 500$ $\Omega$ and $R_\square \sim 750$ $\Omega$ for the (001) and (111) Hall bars respectively, with the corresponding areal charge densities obtained from the low-field Hall coefficient based on a single band model of $n\sim9 \times 10^{13}$/cm$^2$ and $n\sim 2 \times 10^{13}$/cm$^2$ (see Supplementary Information).      

\begin{figure*}[!ht]
\includegraphics[width=18cm]{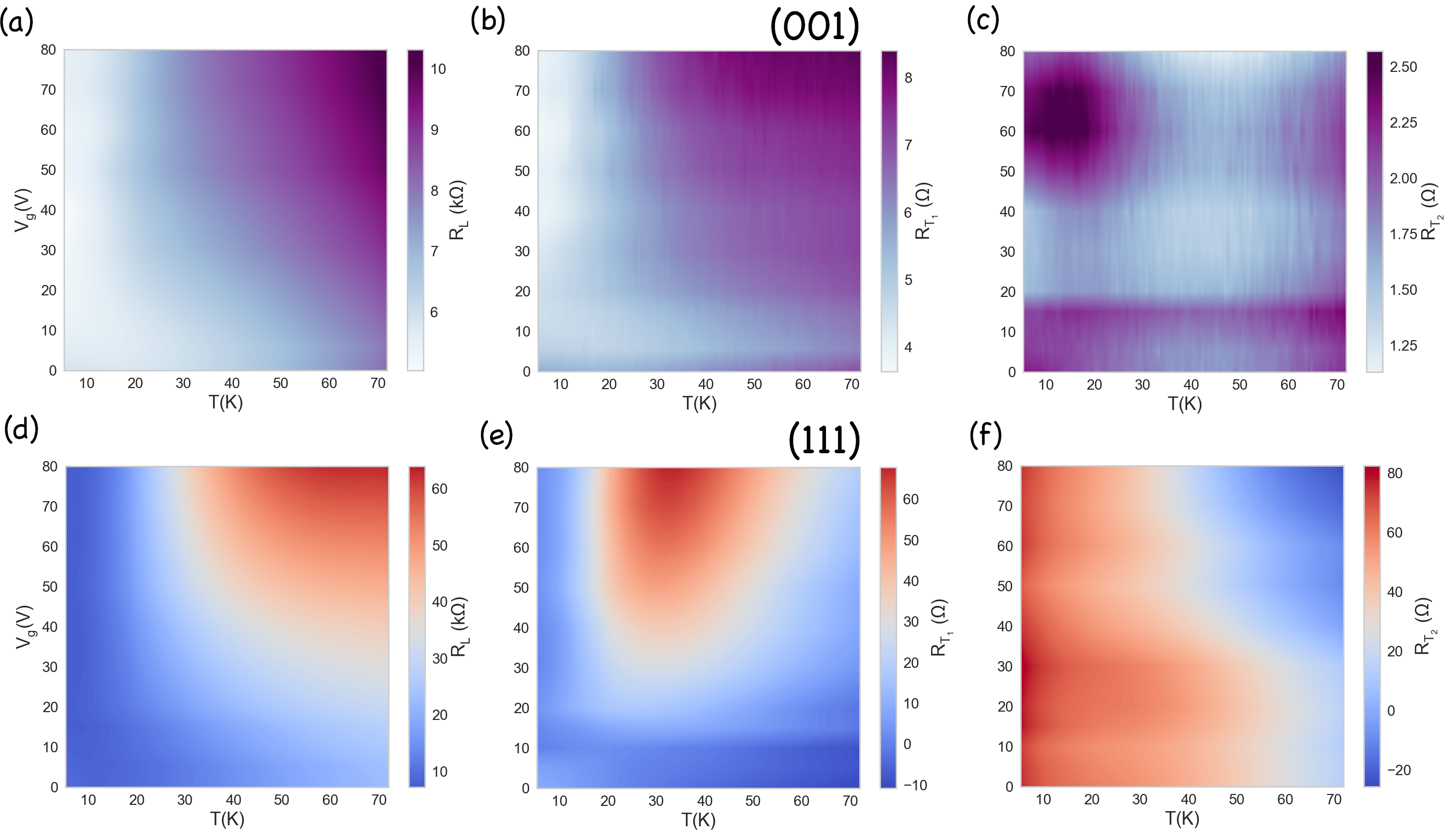}
\caption{  (a)-(c)  Gate voltage and temperature dependence at $B=0$ of the longitudinal resistance $R_L$ and two transverse resistances $R_{T_1}$ and $R_{T_2}$ for a (001) oriented Hall bar in zero external magnetic field.  (d)-(f)  Corresponding data for a (111) oriented Hall bar.  } 
\label{fig:fulldependence}
\end{figure*}

Figures \ref{fig:roft}(c) and (d) show representative longitudinal ($R_L$) and transverse ($R_T$) resistances of Hall bars on (001) and (111) oriented samples as a function of temperature at $V_g$ = 80 V and $B=0$.    A finite $R_T$ is seen in both orientations, with the magnitude of $R_T$ in the (111) Hall bars being much larger in comparison to the (001) Hall bars.  Care was taken to qualify the sample geometry and alignment, as a finite $R_T$ in zero field is most commonly attributed to misalignment of the transverse voltage probes, so that some fraction of $R_L$ contributes to $R_T$.  If so, one would expect $R_T$  to track $R_L$ as a function of temperature.  The data in Figs. \ref{fig:roft}(c) and (d) show that this is not the case for these devices.       

Figure \ref{fig:fulldependence} shows the full temperature and gate voltage dependence of the longitudinal resistance $R_L$ and the two transverse resistances $R_{T_1}$ and $R_{T_2}$ (see Fig. \ref{fig:roft}(a)) for both a (001) and a (111) oriented Hall bar at $B=0$.  There are a number of features of these data that should be noted (these features are reproduced in all the Hall bars we have measured).  First, $R_L$ for both the (001) and the (111) oriented samples has its lowest value at low $T$ and small values of $V_g$, and its highest value at $V_g=80$ V and high $T$.  While $R_L$ at high $T$ and large $V_g$ is about a factor of 6 larger for the (111) sample in comparison to the (001) sample, $R_L$ at low $T$ and small $V_g$ for the two Hall bars is comparable.  Second, the change in $R_{T_1}$,  $R_{T_2}$over the range of $V_g$ and $T$ shown is larger by a factor of 20 or more for the (111) sample in comparison to the (001) sample.  Third, for both (001) and (111) Hall bars, the two transverse resistances $R_{T_1}$ and $R_{T_2}$ measured at points 600 $\mu$m apart on the same Hall bar are qualitatively different from each other in their dependence on $T$ and $V_g$.  This is another indication that the finite zero-field $R_T$ does not arise from misalignment of the transverse voltage probes.  Finally, $R_T$ can change sign as a function of $T$ or $V_g$ for both the (001) and the (111) oriented samples.  An example of this for the (111) sample can be seen in Figs. \ref{fig:fulldependence}(e) and (f). 

\begin{figure}
\includegraphics[width=7cm]{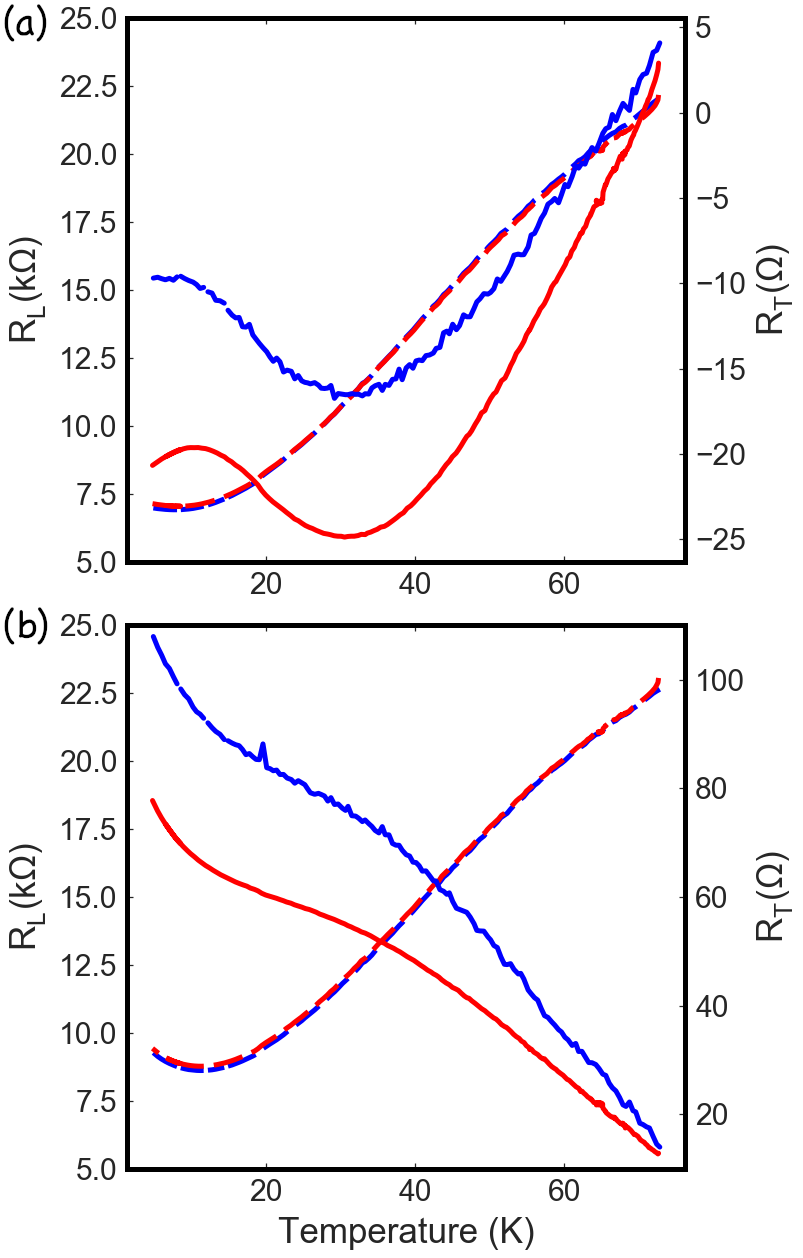}
\caption{ (a) Red and blue solid curves show the transverse resistance $R_T$ measured using the same set of voltage probes with $V_g=0$, $B=0$ for a (111) oriented sample on two different cooldowns.  The dashed curves show the corresponding simultaneously measured longitudinal resistance $R_L$, which lie essentially on top of each other.  (b) Similar data for a different (111) oriented Hall bar.  } 
\label{fig:reproducibility}
\end{figure}

In addition to varying spatially along the length of a single Hall bar,  $R_T$ in zero field also changes quantitatively and qualitatively after warming the sample to room temperature and cooling back down again, even though $R_L$ may not change.  This is shown in Figs. \ref{fig:reproducibility}(a) and (b), which show the simultaneously measured zero-field $R_T$ and $R_L$ for two different (111) oriented devices on two separate cooldowns after warming to room temperature.  In both samples, $R_L$ on the two different cooldowns is essentially unchanged, but the transverse resistance $R_T$ measured on the same voltage probes at low temperature varies by 40\% or more.  More typically, if the devices are left under ambient conditions at room temperature for extended periods of time (i.e., 1-2 weeks), $R_L$ at low temperatures may change by approximately 20\% due to a change in the oxygen vacancy concentration.  In this case, $R_T$ measured using the same probes at low temperatures may change by a larger amount, and even change its qualitative behavior, e.g., it may initially show a peak in resistance at intermediate temperatures for larger $V_g$ as in Fig. \ref{fig:fulldependence}(e), but then show an overall increase or decrease in resistance as in Fig. \ref{fig:fulldependence}(f) on a subsequent cooldown.  The differences in $R_T$ are accentuated below 40 K; at higher temperatures, $R_T$  measured on the same set of probes converges to roughly the same value at higher temperatures, as can be seen from Figs. \ref{fig:reproducibility}(a) and (b).      

\begin{figure*}
\includegraphics[width=16cm]{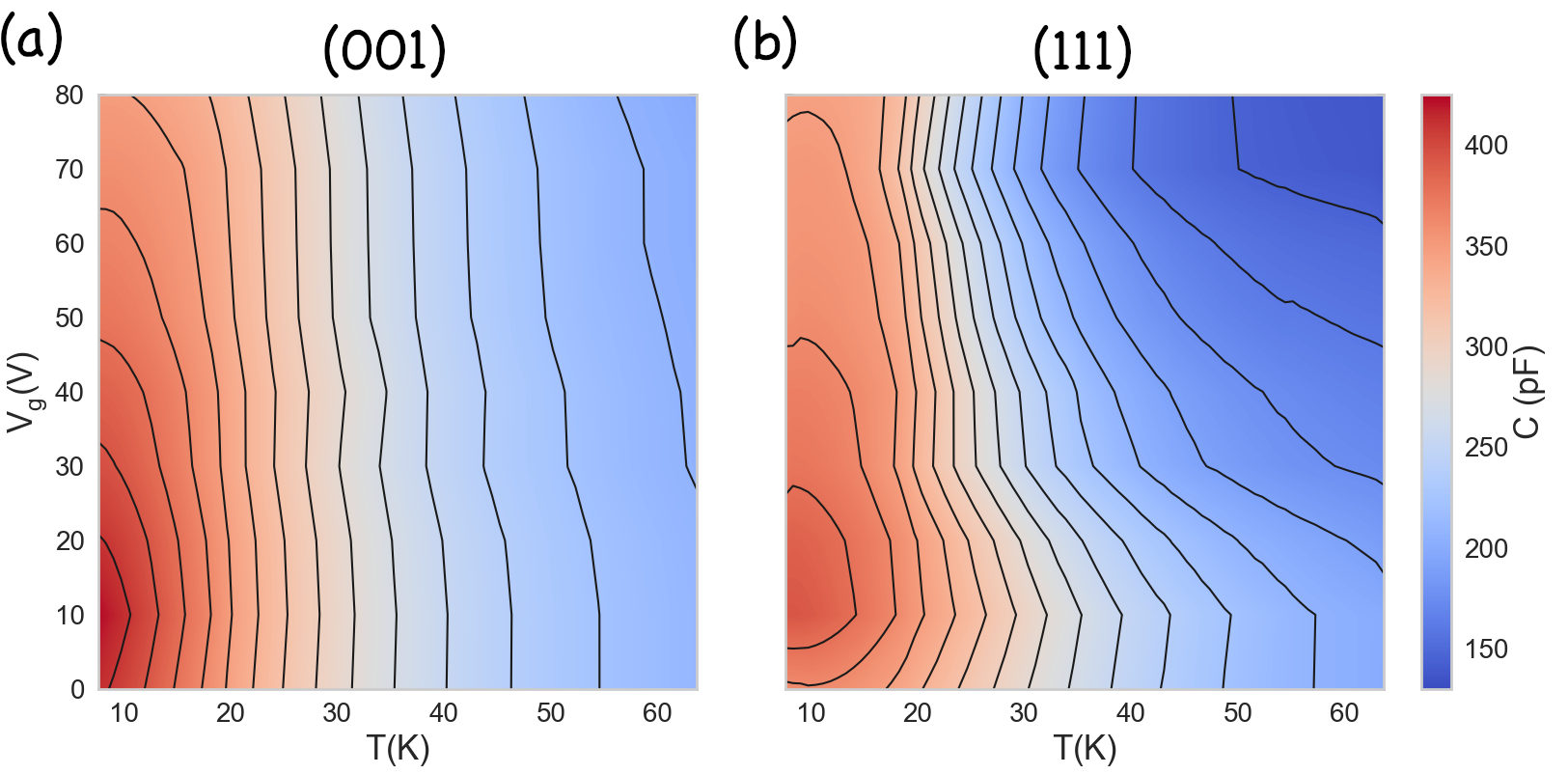}
\caption{\textbf{Temperature and gate voltage dependence of capacitance.}  (a) Capacitance measured between the back gate and a Hall bar on a [001] oriented substrate as a function of $T$ and $V_g$. (b) Similar data for a (111) oriented Hall bar.  } 
\label{fig:capacitance}
\end{figure*}

As previously noted, intrinsic magnetism may give rise to a ZFTR.  The lack of hysteresis in the longitudinal and Hall magnetoresistance indicates that this source, if present does not contribute significantly to the ZFTR (see Supplementary Information).  Recent experimental studies\cite{kalisky,frenkel,frenkel2,goble} have shown that current in STO-based 2DCGs appears to flow preferentially along the domain walls formed between tetragonal domains.   The domain walls form along the \{110\} planes\cite{honig} and range from a few microns to many tens of microns in length.  They are aligned at specific angles to the nominal direction of the injected current in our devices, as shown in Figs.  \ref{fig:roft}(e-g).  For the Hall bars on the  (001) oriented substrates, these angles are $\pm$ 45$^\circ$, while for the Hall bars on the (111) oriented substrates, the angles are $\pm$ 30$^\circ$ and 90$^\circ$ for Hall bars aligned along the $[1\bar{1}0]$ surface crystal direction, and $\pm$ 60$^\circ$ and 0$^\circ$ for Hall bars aligned along the $[\bar{1}\bar{1}2]$ surface crystal direction.  Current flow along these domain walls would give rise to finite off-diagonal components of the resistivity tensor ($\rho_{xy}$, $\rho_{yx}$) that vary as a function of position as they depend on the local domain configuration.  Since the domain configuration is random, and the current is equally likely to be deflected to the right or to the left, the sign of the resulting transverse resistance can be positive or negative, as shown in Figs. \ref{fig:roft}(e-g).  Consequently, for very wide samples encompassing many such domain walls, one might expect the transverse resistance to average to zero, while for narrower devices with only one or a few domain walls, the effect would be larger.  The domain walls are known to reconfigure on thermal cycling and move under the influence of a backgate voltage.\cite{frenkel,frenkel2,ma,honig}  They are also reported to become more polar at temperatures below 40 K.\cite{salje}  These properties are all consistent with the properties of the ZFTR, suggesting that charge transport along domain walls is responsible for the transverse resistance we observe. 

Based on the model above, we would expect the magnitude of the ZFTR in the (001) and (111) oriented samples to be the same.  However, the magnitude of the change in the ZFTR as a function of $T$ and $V_g$ is much larger for the (111) oriented samples in comparison to the (001) oriented samples.  In order to explore the potential reason for this difference, we measured the $T$ and $V_g$ dependence of the capacitance $C$ between the Hall bars and the back gate.\cite{davis}  Modeling the system as a parallel plate capacitor, $C$ directly reflects the dielectric constant $\epsilon$ of the STO.  These capacitance data are shown in Figs. \ref{fig:capacitance}(a) and (b)  for a (001) and a (111) oriented Hall bar respectively.  For both orientations, $C$ increases as $T$ decreases, and at the lowest temperature, decreases with increasing $V_g$, consistent with previous measurements of $\epsilon$ in STO where the electric field was applied along the $<$100$>$ or $<$110$>$ crystal directions.\cite{weaver,muller,muller2}  However, there is a difference in the $V_g$ dependence of $C(T)$ between the two orientations:  For the (001) sample, $C(T)$ is not strongly dependent on $V_g$, while for the (111) sample, there is a significant sharpening of the transition from low capacitance at higher temperatures to higher capacitance at lower temperatures with increasing $V_g$, with the transition region moving from approximately 30 K down to 20 K as $V_g$ increases from 0 to 80 V.  We note that earlier measurements\cite{muller2} of the dielectric constant of STO have been performed with the electric field aligned along the $<100>$ or $<110>$ directions, and no significant differences between different field directions were noted.  In the absence of polar domain walls, it is not immediately clear why the dielectric constant should depend on the direction of the electric field with respect to the crystalline axes, given that the $c$-axes of tetragonal domains are randomly oriented along the $<100>$ directions.  The greater sensitivity of both the ZFTR and the capacitance of our structures when the electric field is aligned along the [111] direction suggests a common origin of dynamic domain walls for both effects.

In summary, we propose that preferential conduction along domain walls is responsible for the zero field transverse resistance observed in our Al$_2$O$_x$ / STO samples. The effect is primarily a function of electric field, temperature, and crystal direction, but is also subject to the dynamics of domain formation in the STO. Capacitance measurements conducted to probe the difference in sample crystal orientation revealed a directional dependence of the dielectric constant, and suggest a common origin around the quantum paraelectric transition. These results have significant implications for making mesoscopic devices from STO-based oxide heterostructures, especially those on [111] substrates.\\                  

\begin{acknowledgements}
We thank A. Balatsky for useful discussions.  This work was supported by the the US Department of Energy, Basic Energy Sciences, under grant number DE-FG02-06ER46346. 
\end{acknowledgements}

\end{document}


<<<<<<< HEAD
\title{Supplementary Information:\\
Observation of zero-field transverse resistance in AlO$_x$/SrTiO$_3$ interface devices }
=======
\title{Supplemental Information:\\
Observation of zero-field transverse resistance at the AlO$_x$ / SrTiO$_3$ interface}
>>>>>>> e808ce844a84c538e6d458287600c06e1e80148b
\author{P.W. Krantz}
\address{Department of Physics, Northwestern University, Evanston, Illinois. 60208, USA}
\author{V. Chandrasekhar}
\address{Department of Physics, Northwestern University, Evanston, Illinois. 60208, USA}
\email{v-chandrasekhar@northwestern.edu}

\date{\today}

\maketitle
\section{Magnetoresistance Investigations} 
<<<<<<< HEAD
\textbf{Investigation of potential magnetism.}
One alternative source of zero-field transverse resistance is intrinsic magnetism. An intrinsic ferromagnetism might give rise to hysteretic magnetoresistance in the longitudinal or transverse resistance as a function of magnetic field through the anomalous Hall effect or anisotropic magnetoresistance. We looked for both of these signatures, and did not find them. Representative traces for simultaneously measured longitudinal and transverse resistances as a function of parallel magnetic field for (111) oriented samples can be found in Figure \ref{fig:MagResComp} (a), while traces for perpendicular field measurements can be found in Figure \ref{fig:MagResComp} (b). Here parallel field refers to magnetic field applied parallel to the plane of the sample. Both field directions show drift in the longitudinal traces upon repeated passes. This is a well known characteristic of STO magnetoresistance traces, and should not be mistaken for hysteretic behavior. In either case, the transverse signal shows no such drift, and is featureless upon repeated field sweeps.
=======
Investigation of potential magnetism.
One alternative source of zero-field transverse resistance is anomalous Hall effect due to magnetic moment or spinful scattering sites. In this case an effective magnetic field in the material results in an offset of the Hall signal in zero applied field, and can change as a function of temperature due to relations with the Fermi level. Often this magnetism leaves signatures on magnetoresistance, either through switching at the level of the coercive field of the innate magnetic moment, or through interactions with magnetic alignment. These interactions show signs in magnetoresistance under perpendicular field (switching, hysteresis, bending of Hall signal), or in parallel field (changes in resistance due to alignment or magnetic domains etc.). We looked for both of these signatures, and did not find them. Representative traces for simultaneously measured longitudinal and transverse resistances as a function of perpendicular magnetic field for (001) and (111) oriented samples can be found in Figure \ref{fig:MagResComp} (a), while traces for parallel field measurements in two perpendicular directions can be found in Figure \ref{fig:MagResComp} (b).
>>>>>>> e808ce844a84c538e6d458287600c06e1e80148b

\begin{figure}[H]
\centering
\includegraphics[width=16cm]{Figures/MagResCompositePaper.png}
<<<<<<< HEAD
\caption{Representative traces of parallel (a) and perpendicular (b) field magnetoresistance showing the $\Delta R/R_0$ of the longitudinal component and the corresponding transverse signals. The parallel field measurements in (a) shows a [110] oriented Hall bar with one of its Hall resistances for fields oriented along the path of the current (here labeled [110] Field) and perpendicular to the path of the current (here labeled [112] Field). The perpendicular field measurement in (b) shows a measurement for the same [110] oriented Hall bar and its two transverse (Hall) resistances.  These two measurements were taken during different cooldowns, resulting in the shift in the zero field Hall signature, as discussed in the main text.}
=======
\caption{Representative traces of perpendicular (a) and parallel (b) field magnetoresistance showing the $\Delta$R/$R_0$ of the longitudinal component and the corresponding transverse signals. The perpendicular field measurement in (a) shows the measurement for the [112] oriented Hall bar and its two transverse (Hall) resistances. The parallel field measurements in (b) shows the same [112] oriented Hall bar with one of its Hall resistances for fields oriented along the path of the current (here labeled parallel) and perpendicular to the path of the current (here labeled perpendicular). Parallel field measurements shown are from a different cooldown than the perpendicular field measurement shown, resulting in the shift in the zero field Hall signature, as discussed in the main text.}
>>>>>>> e808ce844a84c538e6d458287600c06e1e80148b
\label{fig:MagResComp}
\end{figure}

\section{Carrier Density Calculations}
<<<<<<< HEAD
Reported carrier densities were calculated from a single band model $R_H = - 1/ ne$, where for a two dimensional gas, $R_H$ is given simply by the slope of the transverse resistance with magnetic field as shown in the supplementary figure. In all cases, the sign of the slope corresponds to electron-like carriers. 
=======
Reported carrier densities were calculated from a single band model applied to the low temperature magnetoresistance traces described above. We have the Hall coefficient $R_H$ given by $R_H = E_y / (j_x B) = - 1/(ne)$ where $E_y = V_H / width$ is the electric field related to the measured Hall voltage, $j_x$ is the injected current density, $B$ the externally applied magnetic field, $n$ the carrier concentration, and $e$ the charge of the carrier. Because we are in 2D, we substitute for sheet resistance $R_s = V_H / I_x$ where $I_x = j_x * width$ to get the form for our Hall coefficient, $R_H = R_s / B = - 1/ ne$. We measure the slope of the Hall signal, and it is inversely proportional to the carrier concentration. If the slope is negative with $B$ the carriers are predominately electron like, which is the case in our system, and agrees with other measurements on SrTiO$_3$ based devices. 
>>>>>>> e808ce844a84c538e6d458287600c06e1e80148b